\documentclass[twocolumn,showpacs,aps,prl,superscriptaddress]{revtex4}

\usepackage{graphicx}
\usepackage{dcolumn}
\usepackage{amsmath}
\usepackage{epsfig}

\input babarsym

\newcommand{\BABARPubYear}    {04}
\newcommand{\BABARPubNumber}  {043}

\newcommand{\SLACPubNumber} {10903}

\def\figurebox#1#2#3{
    \def\arg{#3}
    \ifx\arg\empty
    {\hfill\vbox{\hsize#2\hrule\hbox to #2{\vrule\hfill\vbox to #1{\hsize#2\vfill}\vrule}\hrule}\hfill}
    \else
    {\hfill\epsfbox{#3}\hfill}
    \fi}

\begin{document}

\begin{flushleft}
\babar-PUB-\BABARPubYear/\BABARPubNumber\\
SLAC-PUB-\SLACPubNumber\\
\end{flushleft}

\title{\large\boldmath
Search for a Charged Partner of the $X(3872)$
in\\ the $B$ Meson Decay $B\to X^{-}K$, $X^{-}\rightarrow J/\psi\pi^-\pi^0$}
%
\author{B.~Aubert}
\author{R.~Barate}
\author{D.~Boutigny}
\author{F.~Couderc}
\author{Y.~Karyotakis}
\author{J.~P.~Lees}
\author{V.~Poireau}
\author{V.~Tisserand}
\author{A.~Zghiche}
\affiliation{Laboratoire de Physique des Particules, F-74941 Annecy-le-Vieux, France }
\author{E.~Grauges-Pous}
\affiliation{Universitad Autonoma de Barcelona, E-08193 Bellaterra, Barcelona, Spain }
% ...
\collaboration{The \babar\ Collaboration}
\noaffiliation

\date{\today}

\begin{abstract}

We search for a charged partner of the $X(3872)$ in the decay
$B\rightarrow X^- K$, $X^- \rightarrow J/\psi \pi^- \pi^0$,
using {234} million $B\kern 0.18em\overline{\kern -0.18em B}$ events
collected at the $\Upsilon(4S)$ resonance with the
$\mbox{\slshape B\kern-0.1em{\smaller A}\kern-0.1em B\kern-0.1em{\smaller A\kern-0.2em R}}$
detector at the PEP-II $e^+ e^-$ asymmetric-energy storage ring.
The resulting product branching fraction upper limits are
$\cal B$$(B^0\rightarrow X^{-}K^+$, $X^-\rightarrow J/\psi\pi^-\pi^0)$
$< 5.4 \times 10^{-6}$
and
$\cal B$$(B^-\to X^{-}{\kern 0.2em\overline{\kern -0.2em K}}^0$, 
$X^- \rightarrow J/\psi\pi^-\pi^0)$
$< 22 \times 10^{-6}$ at the $90\%$ confidence level. 

\end{abstract}

\pacs{13.25.Hw, 14.40.Gx, 12.39.Mk}

\maketitle

\def\Xpm {\ensuremath{X^\pm}\xspace}

\def\sigcut{$|m(\jpsi\pipm\piz)-3872 \mevcc|<12\mevcc$}
\def\sidcut{$48<|m(\jpsi\pipm\piz)-3872 \mevcc|<72\mevcc$}
\def\bpmmode{$\Bm\to\jpsi\pim\piz\KS$}
\def\b0bmode{$\Bz\to\jpsi\pim\piz\Kp$}

\indent \indent The
discovery of the $X(3872)$ by the
Belle Collaboration~\cite{x3872-belle},
has been confirmed by
the CDF~\cite{x3872-cdf}, D0~\cite{x3872-d0} and
$\babar$~\cite{x3872-babar} Collaborations.
Numerous theoretical explanations
have been proposed for
this high-mass, narrow-width state
decaying into $\jpsi\pi^+ \pi^-$.
The possibilities~\cite{theory-general}
include
a bound state of $\Dstar\Db$ very close the $\Dstarz\Dzb$
threshold~\cite{theory-molecule},
a hybrid charmonium state~\cite{theory-hybrid},
a diquark-antidiquark state~\cite{theory-diquark},
and a conventional charmonium state~\cite{theory-charmonium}.
The Cornell potential model~\cite{eichten}
predicts
a
$^3D_2$
($J^{PC}=2^{--}$)
charmonium state with a $3.830\gevcc$ mass.
This state is expected to be very narrow
since the decay to $D\Dbar$ is forbidden by parity
and could decay into an isoscalar
$\jpsi\pi^+ \pi^-$ final state.

The charmonium state, however, should also have a significant branching
ratio for the radiative decay to $\gamma\chi_{c1}$~\cite{eichten}, which
was not observed for the $X(3872)$ by Belle~\cite{x3872-belle}.
A more detailed examination of the $X(3872)$
observed by Belle~\cite{x3872-belle}
and $\babar$~\cite{x3872-babar} indicates that
the $\pi^+ \pi^-$ mass
distributions peak
near the kinematic upper limit and are
consistent with the decay $\rho^0 \to\pi^+ \pi^-$.
However, due to limited statistics a spin-parity analysis
has not been performed.
If the observed decay is $X(3872)\to\jpsi\rho^0$ and
if these states and
their decays respect isospin symmetry,
then there must be a $X(3872)^-$,
which decays to $\jpsi\rho^-$, and the rate
for $B\to\X^-K$ should be
twice that for $B\to\X^0K$. This would
make experimental detection of
the $X^-$ quite favorable.
To test this hypothesis, we have performed
a search for the $B$-meson decays,
$\Bz \to X^-\Kp$
and
$\Bm \to X^-\KS$,
where $X^-\to\jpsi\pim\piz$~\cite{chargeConj}.

\indent \indent
Data were collected at
the PEP-II asymmetric-energy $e^{+}e^{-}$ storage ring with the
$\babar$ detector, which is described in detail elsewhere~\cite{babar-det}. 
The data used in this analysis correspond to a total integrated luminosity of 
$212$ fb$^{-1}$ taken on the $\FourS$ resonance, producing a sample 
of $234.4\pm 2.6$ million \BB events ($N_{\BB}$).
The $\babar$ detector uses a silicon vertex tracker (SVT) and a 
40-layer drift chamber (DCH), both in a 1.5-T solenoidal magnetic field 
to detect charged particles and measure their momenta and
energy loss ($dE/dx$). Photons, electrons, and neutral hadrons
are detected in a
CsI(Tl)-crystal electromagnetic calorimeter (EMC). 
An internally reflecting ring-imaging
Cherenkov detector (DIRC) provides 
particle
identification 
information that is complementary to that from $dE/dx$.
Penetrating
muons and neutral hadrons are identified  
by resistive-plate chambers 
in the steel flux return (IFR).
Track-selection criteria in this analysis follow previous
$\babar$ analyses~\cite{babar-charmonium}.

\indent \indent This analysis commences with charged and neutral candidate
selections.
Each charged-track candidate is required to be detected in at least 12 DCH 
layers and to have a transverse momentum greater than $100\mevc$. 
If it is not associated with a $K^0_{S}$ decay, the candidate must 
extrapolate to a point near the collision axis.

A charged kaon or pion candidate is selected on the basis of $dE/dx$ 
information from the SVT and DCH, and the Cherenkov angle measured by the DIRC.
An electron candidate is required to have a good match 
between the expected and measured $dE/dx$ in the DCH, and 
the Cherenkov angle in the DIRC. 
The ratio of the shower energy measured in the EMC to the momentum
measured in the DCH, and the number of 
EMC crystals associated with the track
candidate must be appropriate for an electron. 
A muon is  selected on the basis of energy deposited in 
the EMC, the number and distribution of hits in the IFR, and 
the match between the IFR hits and the extrapolation of 
the DCH track into the IFR. A more detailed explanation of particle
identification  (PID) is given elsewhere~\cite{babar-charmonium,kaon-pid}.

A photon candidate is identified from energy deposited in contiguous 
EMC crystals, summed to form a cluster that has 
total energy greater than 30\mev and a shower shape consistent 
with that expected for an electromagnetic shower. 

The decay modes we use to identify
$\Bz\to\jpsi\pim\piz\Kp$
and $\Bm\to\jpsi\pim\piz\KS$ are $\jpsi\to e^+ e^-$, 
$\jpsi\to \mu^+  \mu^-$, 
$\pi^0 \to \gamma\gamma$, 
and $K_{S}^{0}\to \pi ^{+}\pi ^{-}$. 
They are selected to be within the mass 
intervals 
$2.95<m(e^{+}e^{-})<3.14\gevcc$, 
$3.06<m(\mu^{+}\mu^{-})<3.14\gevcc$, 
$0.119<m(\gamma\gamma)<0.151\gevcc$, 
and 
$0.4917<m\left( \pi ^{+}\pi^{-}\right)<0.5037$ $\gevcc$.
We take a larger mass interval for $e^+e^-$ than for 
$\mu^+\mu^-$ to accept
events in which part of the energy is
carried away by 
bremsstrahlung photons.
The orientation of the displacement vector
between the $K_{S}^0$ decay vertex 
and
the \jpsi vertex 
in the lab frame is required to be 
consistent with the $K_{S}^0$ momentum direction.

The search for $B$ signal events  
utilizes two kinematic variables: 
the energy difference $\Delta E$ between the energy of the $B$ candidate and 
the beam energy $E_{\rm{b}}^{*}$ in the $\FourS$ rest frame, and the beam-energy-substituted 
mass $\mes \equiv \sqrt{\left( E_{\rm{b}}^{*}\right)^2 -\left( p_{\rm{B}}^{*}\right) ^{2}}$, where 
$ p_{\rm{B}}^{*}$ is the reconstructed momentum of the $B$ candidate in the $\FourS$ frame.
Signal events should have 
$\mes \approx m_{B}$, where $m_{B}$ is the mass of the
$B$-meson~\cite{pdg}, and $|\Delta E|\approx 0$. 

Before the data were analyzed, the selection criteria were optimized 
and fixed
separately for the charged and neutral $B$ mode using 
a Monte Carlo (MC) simulation of signal and known backgrounds.
The number of reconstructed MC signal events $n_{\rm{s}}^{\rm{mc}}$ and the number of
reconstructed MC background events $n_{\rm{b}}^{\rm{mc}}$
(scaled to the integrated luminosity) were used 
to estimate the sensitivity ratio
$n_{\rm{s}}^{\rm{mc}} / (a/2 + \sqrt{n_{\rm{b}}^{\rm{mc}}})$~\cite{punzi}, 
where $a$, the number of standard deviations of significance
desired, was set to 3.
Note that the maximum of this ratio is
independent of the unknown signal
branching fraction.
This ratio was maximized by
varying the selection criteria
on  
$\Delta E$,
$\mes$, 
the $X^-(\jpsi\pim\piz)$ mass,
the $K_{S}^{0}(\pi^+ \pi^-)$ mass,
the $K_{S}^{0}$ decay-length significance,
the $\gamma\gamma$ invariant mass,
and the
particle-identification criteria for electrons, muons
and charged kaons.
When there was more than one candidate per event after applying the 
optimized cuts (on average there were 1.3 candidates/event), the candidate
with the smallest value of $|\DeltaE|$
was chosen. The selections $|\mes-m_{B}| < 5$ $\mevcc$
and $|\Delta E|< 20\mev$ (signal-box region) were 
found to be optimal for selecting signal events.
The plots that follow include only one candidate per event, except
for the plots showing \DeltaE itself.

\begin{figure}[h]
\begin{center}
\includegraphics[width=0.45\textwidth]{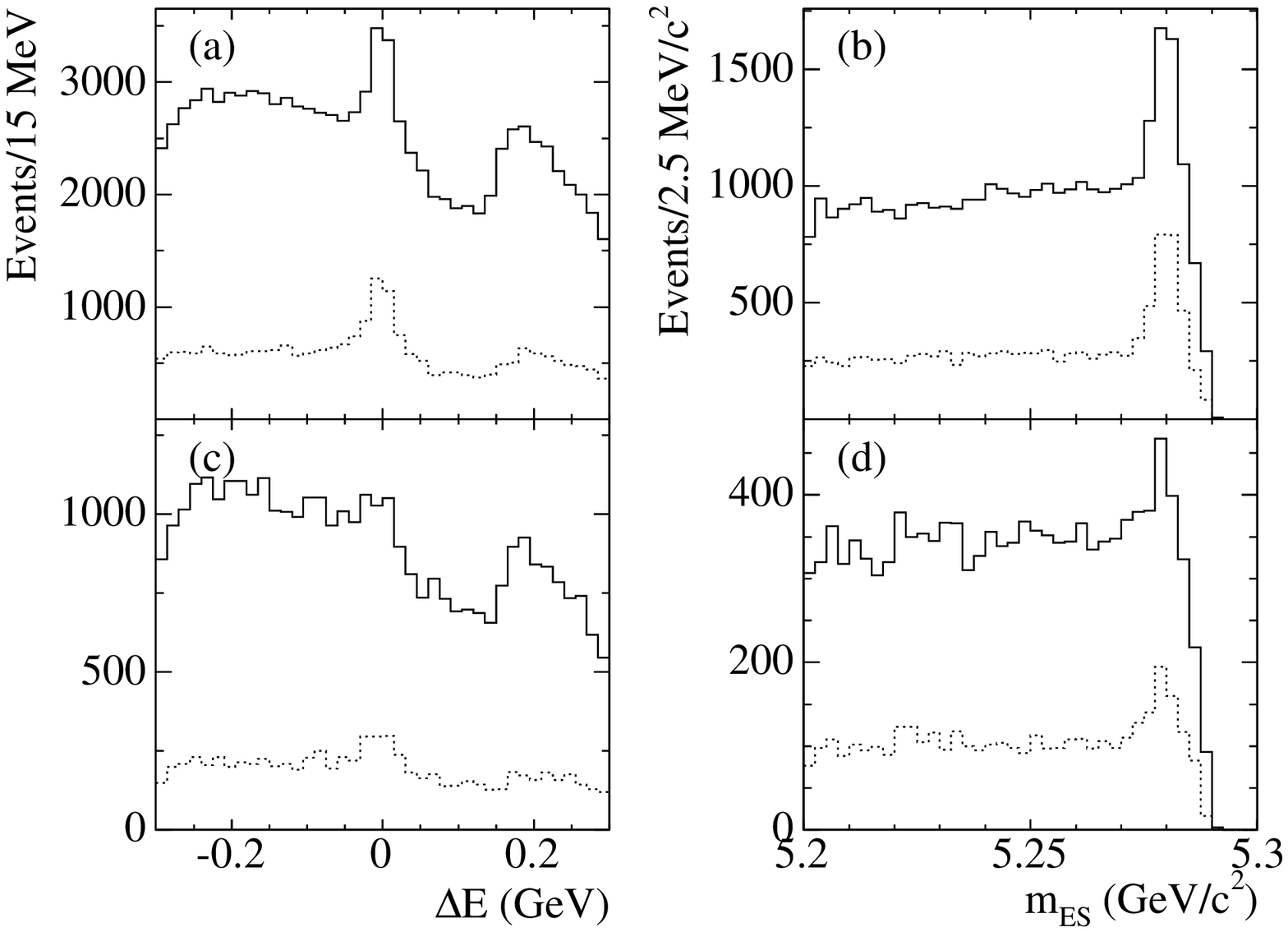}
  \caption{\label{fig:mes+deltae}
  The $\DeltaE$ (a) and  $\mes$ (b) distributions for the
\b0bmode \ mode
  and the $\DeltaE$ (c) and  $\mes$ (d) distributions for the
\bpmmode \ mode using the optimized cuts. The dotted line shows the same
with the additional cut $0.67<m(\pim\piz)<0.87\gevcc$.}
\end{center}
\end{figure}

The $\Delta E$ and $\mes$ distributions for
the neutral and charged $B$ modes
after we apply all the optimized cuts, except the cut for the 
variable plotted, are shown in Figs.~\ref{fig:mes+deltae} (a-d).
A clear peak is observed at zero in the $\Delta E$
distribution and near 5.279 $\gevcc$ in the $\mes$
distribution. 
The other feature in the $\Delta E$ plots is
a wide peak near 0.2 $\gev$ which is due
to $B \to\jpsi K^*$ decays combined
with a random pion.

\begin{figure}[h]
\begin{center}
\includegraphics[width=0.45\textwidth]{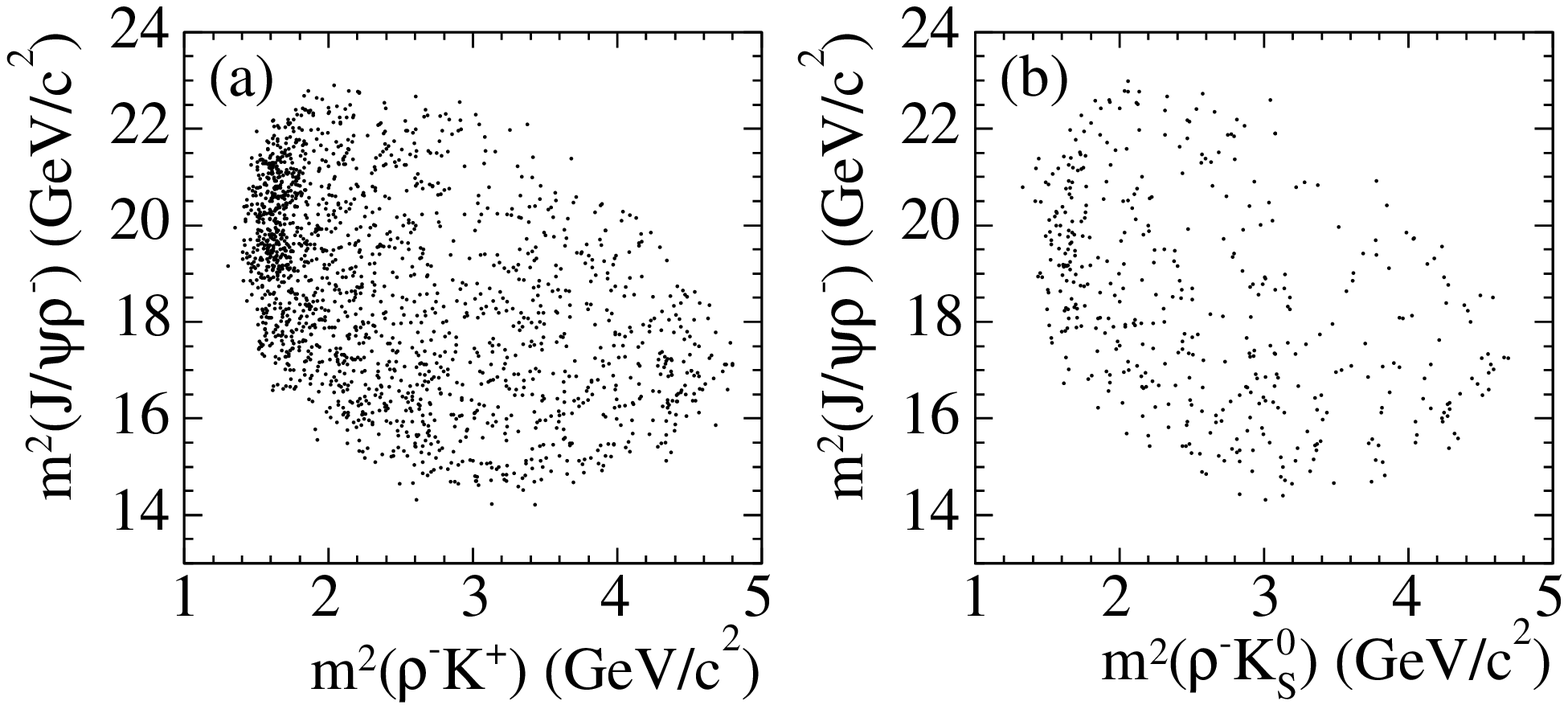}
  \caption{\label{fig:dalitz}
  The $m^2(\jpsi\rho^-)$ versus the  
   $m^2(\rho^-\Kp)$ 
distributions (a) for \b0bmode \
and the
   $m^2(\jpsi\rho^-)$ versus the  
   $m^2(\rho^-\KS)$
distributions (b) for \bpmmode. A $B\to\jpsi K_1$ signal can be seen, however
there is no indication for an enhancement in the $\jpsi\rho^{-}$ mass spectrum.
  }
\end{center}
\end{figure}

The Dalitz plots in Fig.~\ref{fig:dalitz} for the charged- and neutral-$B$
modes use events in the signal-box region and include a mass cut of 
$0.67<m(\pim\piz)<0.78$ $\gevcc$ to select the $\rho^-$ mass region.
There are clear bands for
$K_1^0(1270) \to\Kp\rho^-$
and
$K_1^-(1270) \to\KS\rho^-$
corresponding to the decays
$\Bm\to\jpsi K_1^-$
and
$\Bz\to\jpsi K_1^0$
previously observed by Belle~\cite{belle-k1}.

\begin{figure}[h]
\begin{center}
\includegraphics[width=0.45\textwidth]{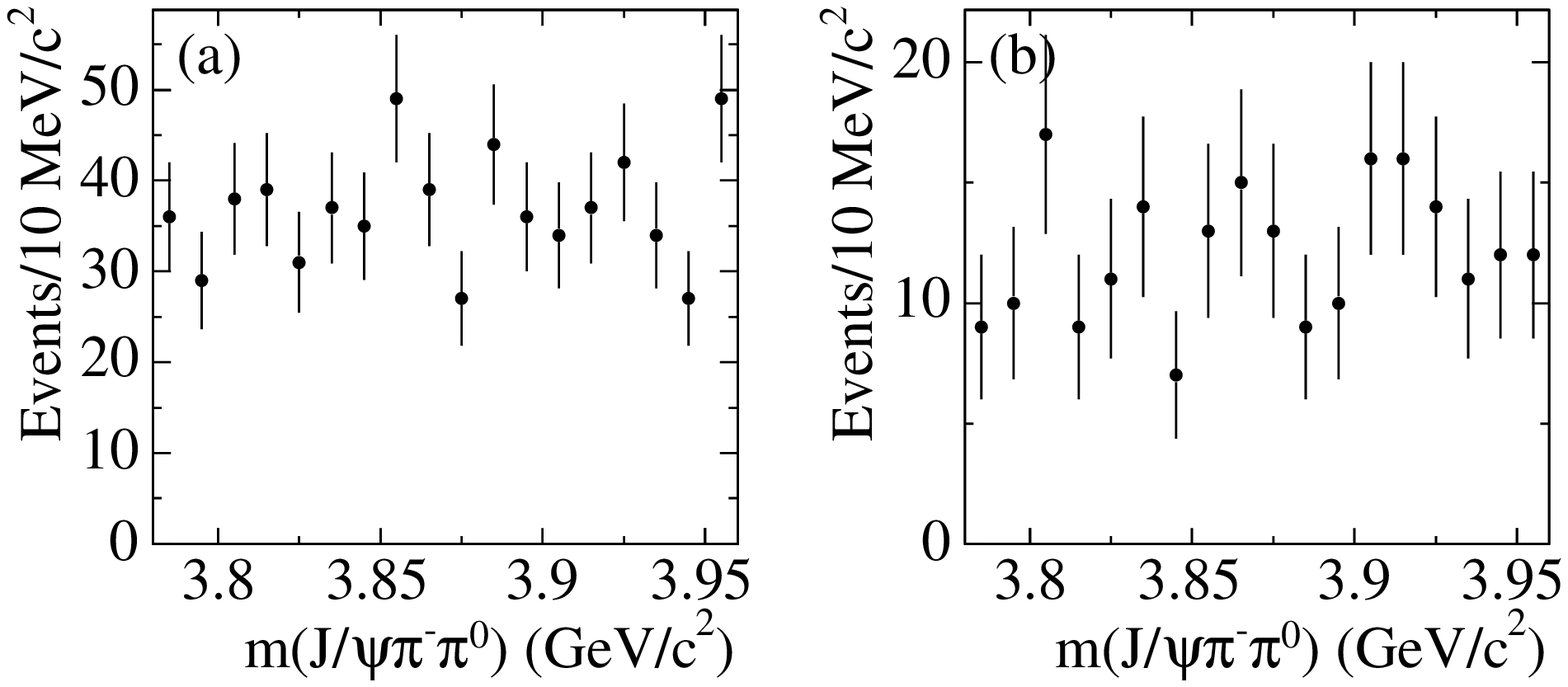}
  \caption{\label{fig:xmass}
    The $\jpsi\pim\piz$ invariant mass
    in 10 $\mevcc$ bins for (a) \b0bmode and (b) for \bpmmode.
    No indication for the decay $X^{-}\to\jpsi\pim\piz$ can be found.
  }
\end{center}
\end{figure}

The $\jpsi\pim\piz$ mass spectra  
from the neutral and charged $B$ modes
are shown in Fig.~\ref{fig:xmass} without a $\rho$ mass cut.
No charged signal, $X^- \to\jpsi\pim\piz$,
is evident at $3.872\gevcc$.

Extracting an upper limit for $X^-\to\jpsi\pim\piz$ requires
examining the $\jpsi\pim\piz$
mass, \mes, and \DeltaE distributions.  A signal from $B\to X^-K$,
$X^-\to\jpsi\pim\piz$
should produce signal peaks in all three distributions. 
Background from $B\to\jpsi\pim\piz K$ in which the $\jpsi\pim\piz$
is nonresonant would produce
peaks in the \mes and \DeltaE distributions but have a
flat $\jpsi\pim\piz$ mass distribution near 3.872\gevcc.
The combinatoric background will not create peaks in any of the three
distributions and should produce an \mes distribution whose shape can be
parametrized by an ARGUS function~\cite{argus}. To estimate the number of signal events 
$(n_{S})$, we count the number of observed events $(n_{obs})$ in the signal
region and subtract the estimated number of combinatoric background events
$(n_{comb})$ and the estimated number of peaking background events $(n_{peak})$. 

We obtain $n_{obs}$ by counting the number
of events satisfying $\left|\mes-m_{B}\right|<5\mevcc$, $\left|
\DeltaE\right|<20\mev$, and $\left|m(\jpsi\pim\piz)-
3872\right|<12\mevcc$. We extract $n_{comb}$ from
the \mes distribution obtained after requiring  $\left|\DeltaE\right|
<20\mev$, and $\left|m\left(\jpsi\pim\piz
\right)-3872\right|<12\mevcc$. 
These \mes distributions for the neutral and charged $B$ modes are
separately fit with the sum of a signal Gaussian function and an ARGUS
function. 
The resulting ARGUS function
is integrated over the \mes range,  $\left|\mes-m_{B}\right|<5\mevcc$,
to produce $n_{comb}$. The error $\sigma _{comb}$ is obtained from
the fit error on the normalization of the ARGUS function. The resulting
values for $n_{comb}$ and $\sigma _{comb}$  are listed in Table 1.
\begin{table} [h]
\caption{
Efficiencies, number of signal-box events,
and estimated number of background events $n_b$
($n_{peak}+n_{comb}$) for the neutral and charged $B$ decays.
}
\begin{tabular}{lccccc}
\hline\hline\\[-0.2cm]

Mode & $\epsilon$ & $n_{obs}$ & $n_{peak}\pm \sigma _{peak}$
& $n_{comb}\pm \sigma _{comb}$ & $n_{b}\pm \sigma _{b}$    \\
\hline \\[-0.2cm]

$\Bz$ &   $10.65\% $&96 & $35.2\pm 8.4$ &$77.6\pm6.6$
&$112.8\pm 10.7$  \\
$\Bm$ & ${8.50\%}$&36 & $ 2.0\pm 5.0$ &$29.3\pm4.1$
&$31.3\pm 6.5$ \\
\hline
\end{tabular}
\label{table-upperlimit-bf}
\end{table}

We extract $n_{peak}$ from the
\mes distribution obtained after requiring $\left|\DeltaE\right|<20\mev$,
and $48<\left|m\left(\jpsi\pim\piz\right)
-3872\right|<72\mevcc$ which is twice the
mass range of the signal band. These \mes distributions for the neutral-
and charged-$B$ modes are separately fit with the sum of a Gaussian
function and an ARGUS function. 
We calculate $n_{peak}$ by counting the number of events in the
$\left|\mes-m_{B}\right|<5\mevcc$ region, subtracting the number
of combinatoric events obtained from integrating the ARGUS function over the
same range, $\left|\mes-m_{B}\right| <5\mevcc$, and finally
dividing the result by two.  Note that the Gaussian distribution
used in all fits has a width fixed to
the value determined from a fit to the \mes distribution obtained
using both the $\jpsi\pim\piz$ signal band and the $\jpsi\pim\piz$
sideband. The error $\sigma _{peak}$ is obtained by adding
in quadrature the Poisson errors on the number of events in $\left|
\mes-m_{B}\right| <5\mevcc$\ and the fit errors on the normalization
of the ARGUS\ function. The resulting values for $n_{peak}$ and $\sigma
_{peak}$ are listed in Table~\ref{table-upperlimit-bf}.

The total background $n_b$ is the sum of the
peaking and combinatoric backgrounds and its error $\sigma_b$
combines in quadrature
the errors from the peaking and combinatoric backgrounds.
The backgrounds and their errors are summarized in 
Table~\ref{table-upperlimit-bf}.

The efficiencies $\epsilon$ for the processes,
$\Bz\to X^-\Kp$,
$X^-\to\jpsi\pim\piz$
and
$\Bm\to\X^-\KS$,
$X^-\to\jpsi\pim\piz$
are determined by MC simulation
with an $X^-$ signal of
zero width, mass 3.872 $\gevcc$, and a model
consisting of the
sequential isotropic two-body decays
$B\to X^-K$, $X^-\to\jpsi\rho^-$
and $\rho^- \to \pim\piz$.

These efficiencies are corrected to account for the small differences
observed in PID, neutral-particle detection, and tracking efficiency that
are found by comparing well-understood control samples taken from data
and MC. The final efficiencies for each mode are listed
in Table~\ref{table-upperlimit-bf}. 

The systematic errors include uncertainties
in the 
number of \BB events in the data sample,
secondary branching fractions,
efficiency calculation due to limited MC statistics,
decay model for the generated events,
background parametrization, PID,
charged particle tracking,
and \piz reconstruction.
The individual uncertainties 
are given as percentages in Table~\ref{table-sys}.
The secondary branching fractions~\cite{pdg} include
$\BR(\jpsi\to e^+e^-,~\mu^+\mu^-)=0.1181\pm 0.0014$ and
$\BR(\KS\to \pip\pim)=0.686\pm0.0027$.
The decay-model uncertainty is estimated
by comparing the efficiencies for
phase space 
and
different decay models~\cite{pakvasa-suzuki}
with
$J^{PC}=1^{++}$ 
and
$J^{PC}=2^{--}$. 
\begin{table} [h]
\caption{
Percentage systematic errors on the branching ratios from the 
neutral and charged $B$ decay modes.
}
\begin{tabular}{lcc}
\hline\hline
Systematic Errors(\%) & \Bz &  \Bm \\ \hline
$N_{\BB}$ & 1.1 & 1.1 \\ 
Branching fractions & 5.3 & 5.3 \\ 
MC statistics & 2.1 & 2.3 \\ 
MC decay model & 1.1 & 3.0 \\ 
Bkgd sideband width & 0.3 & 1.7 \\ 
Particle ID & 5.0 & 5.0 \\ 
Tracking $\pim$ & 1.4 & 1.4 \\ 
Tracking $\Kp$ & 1.4 & - \\ 
Tracking $K_{S}^{0} \to \pi^+\pi^-$ & - & 2.6 \\ 
Tracking $\jpsi \to e^+e^-, \mu^+\mu^-$ & 1.8 & 1.8 \\ 
$\pi ^{0}$ reconstruction efficiency & 3.2 & 3.2 \\ \hline
TOTAL ($\sigma_{sys}$) & 8.8 & 9.7 \\ \hline
\end{tabular}
\label{table-sys}
\end{table}
The background parametrization uncertainty is estimated by
varying the background sideband width, 
refitting the
$\mes$ distributions, and recalculating the
number of events.
The uncertainties in PID, charged-tracking
efficiency, and \piz-reconstruction
efficiency are determined by studying
control samples~\cite{babar-charmonium}.
The total fractional errors $\sigma_{sys}$, listed at the bottom of
Table~\ref{table-sys},
are determined by adding the 
individual contributions
in quadrature.

The probability distribution of the signal events is modeled
as a Gaussian with a mean $n_s$ and standard deviation $\sigma_s$.
For each $B$-decay mode the mean is $n_s$ = $n_{obs}-n_b$ and the
sigma is $\sigma_s$ = $\sqrt{n_{obs}+\sigma_b^2+n_s^2\sigma_{sys}^2}$.
The systematic error is added in quadrature and scales the
errors on $n_{obs}$ and $n_b$ by the same fraction.
We note the mean values $n_s$, for the charged and
neutral modes are consistent with zero, within errors.

The number of events $N_{90}$ corresponding to the 90$\%$ confidence
level (C.L.) upper limit is calculated using
the Gaussian probability distribution
with the assumption that the number of signal events is
always greater than zero.
The integral of the distribution from zero
to $N_{90}$ will be 90$\%$ of the total area above zero.
Combining 
$N_{90}$, 
$\epsilon$, $N_{\BB}$, and the secondary branching fractions, 
we obtain $90\%$ C.L. upper limits for the neutral and charged $B$ modes of
$<5.4\times10^{-6}$ 
and
$<11\times10^{-6}$, respectively.
For completeness we include the 
central value ($68\%$ confidence interval) 
for the
branching fraction using
the $n_s \pm \sigma_s$ values. 
The neutral and charged $B$ mode branching
fractions are
   $(-5.7 \pm 4.9) \times10^{-6}$
and
   $(2.0\pm 3.8)\times10^{-6}$,
respectively.
The results are summarized in
Table~\ref{table-3}.
\begin{table}[h] 
\caption{The estimated number of signal events, 
90\% C.L. upper limit of signal events, 
the branching fraction upper limits,
and the branching fraction $\BR$ for the
decay modes $\Bz\to X^{-}\Kp$ and $\Bm\to X^{-}\KS$.
}
\begin{tabular}{lcccc}
\hline\hline\\[-0.2cm]

Mode &  $n_{s}\pm \sigma _{s}$ &$N_{90}$& $90\%$ C.L. $(\times 10^{-6})$ & $\BR (\times 10^{-6})$    \\ 
\hline \\[-0.2cm]

$\Bz$ &$-16.8\pm 14.7$ & $15.9$ & $< 5.4$ & $-5.7\pm4.9$\\ 
$\Bm$ &$4.7\pm 8.8$&$17.8 $& $< 11$ & $2.0\pm3.8$ \\ 
\hline 
\end{tabular}
\label{table-3}
\end{table}

\indent \indent We test the isovector-$X$ hypothesis 
at a mass of $3872\mevcc$
using a likelihood ratio test~\cite{pdg}.
Here we determine the ratio of the
two probabilities from the
null $(H_0)$
and 
signal $(H_1)$ 
hypotheses using our experimental 
observation of $96$ events in the
signal-box.

The null hypothesis assumes the
background produced all the observed signal-box events.
Assuming the background probability distribution 
is a Gaussian function with mean $n_b$ and width $\sigma_b$,
we calculate a  probability
of $P(H_0)$=$5.82\times 10^{-2}$
to measure $96$ or fewer events.

The isovector signal hypothesis 
predicts the product
branching fractions 
to be related by
$\BR (B\to X^- K,$ $X^-\to\jpsi\rho^-)=2\times\BR (B\to X(3872) K,$ $X(3872) \to\jpsi\rho^0)$.
Using the $\babar$ branching fraction~\cite{x3872-babar}
$\BR (\Bm \to X(3872)\Km,$ 
$X(3872) \to\jpsi\pip\pim)$
$=(1.28\pm 0.41) \times 10^{-5}$
and assuming all $\pip\pim$ decays originate
from $\rho^0$, we
expect  
$\BR (\Bz\to X^-\Kp,$ 
$X^- \to\jpsi\rho^-)$
$=(2.56\pm 0.82) \times 10^{-5}$.
This would produce 
$75\pm 25$ observed signal events
in a data sample of $234$ million \BB events.
The error combines the uncertainty on the branching 
fraction and the systematic error $\sigma_{sys}$ on our
efficiency.
The probability distributions for the
signal events and the estimated background
events are modeled as two uncorrelated Gaussian functions.
The probability of observing $96$ or fewer events (including background)
with this probability distribution 
is $P(H_1)$=$8.41 \times 10 ^{-5}$.

The likelihood ratio $(\lambda)$ test of the null hypothesis relative to
the signal hypothesis yields $\lambda$ = $P(H_0) / P(H_1)$ = $692$. 
This corresponds
to a probability of less than 1 part in 600 that the isovector-$X$ hypothesis
is compatible
with the outcome of our search for $\Bz\to X^-\Kp,$ $X^-\to\jpsi\pim\piz$.
Performing the same study in our search for $\Bm\to X^-\KS,$ $X^-\to\jpsi\pim\piz$
we obtain $\lambda=17$. The combined likelihood ratio is $1.1\times 10^4$.
Our result does not support the hypothesis that the $X(3872)$ is an
isovector particle decaying to $\jpsi\rho$.

\indent\indent
In conclusion, we have performed a 
search for a charged
partner of the $X(3872)$ decaying to 
$\jpsi\pim\piz$. Our results set 
upper limits on
the
product branching
fractions of
$\BR(\Bz\to X^-\Kp,$ $X^-\to\jpsi\pim\piz)< 5.4 \times 10^{-6}$
and
$\BR(\Bm\to X^-\Kzb,$ $X^-\to\jpsi\pim\piz)=
2\times\BR(\Bm\to X^-\KS,$ $X^-\to\jpsi\pim\piz)<22\times 10^{-6}$
at the $90\%$ confidence level.

We exclude the isovector-$X$ hypothesis
with a likelihood ratio test which favors the null hypothesis by a factor
$1.1\times 10^4$ over the isovector signal hypothesis.

\indent \indent
We are grateful for the 
extraordinary contributions of our \pep2\ colleagues in
achieving the excellent luminosity and machine conditions
that have made this work possible.
The success of this project also relies critically on the 
expertise and dedication of the computing organizations that 
support \babar.
The collaborating institutions wish to thank 
SLAC for its support and the kind hospitality extended to them. 
This work is supported by the
US Department of Energy
and National Science Foundation, the
Natural Sciences and Engineering Research Council (Canada),
Institute of High Energy Physics (China), the
Commissariat \`a l'Energie Atomique and
Institut National de Physique Nucl\'eaire et de Physique des Particules
(France), the
Bundesministerium f\"ur Bildung und Forschung and
Deutsche Forschungsgemeinschaft
(Germany), the
Istituto Nazionale di Fisica Nucleare (Italy),
the Foundation for Fundamental Research on Matter (The Netherlands),
the Research Council of Norway, the
Ministry of Science and Technology of the Russian Federation, and the
Particle Physics and Astronomy Research Council (United Kingdom). 
Individuals have received support from 
CONACyT (Mexico),
the A. P. Sloan Foundation, 
the Research Corporation,
and the Alexander von Humboldt Foundation.


\begin{thebibliography}{99}
\bibitem{chargeConj}
Charge conjugate reactions are included implicitly throughout.
\bibitem{punzi}
G. Punzi, ``Sensitivity of searches for new signals and its optimization,"
eprint physics/0308063
\end{thebibliography}
\end{document}